\documentclass[prl,twocolumn,showpacs,superscriptaddress,preprintnumbers,amssymb]{revtex4-2}
\usepackage{graphicx}
\usepackage{dcolumn}
\usepackage{bm}
\usepackage[nointegrals]{wasysym}
\usepackage{color}
\usepackage{physics}
\usepackage{hyperref}

\newcommand{\beq}{\begin{equation}}
\newcommand{\eeq}{\end{equation}}
\newcommand{\beqn}{\begin{eqnarray}}
\newcommand{\eeqn}{\end{eqnarray}}

\newcommand{\ra}{\rightarrow}

\newcommand{\cF}{ {\cal F} }

\newcommand{\cH}{ {\cal H} }

\newcommand{\cZ}{ {\cal Z} }

\newcommand{\U}{\mathrm{U}}

\newcommand{\cx}[1]{{\color{black} #1}}

\begin{document}

\title{Decoherence through Ancilla Anyon Reservoirs}

\author{Nayan Myerson-Jain}

\affiliation{Department of Physics, University of California,
Santa Barbara, CA 93106}

\author{Taylor L. Hughes}

\affiliation{Department of Physics and Institute for Condensed Matter Theory, University of Illinois at Urbana-Champaign,
Urbana, IL 61801}

\author{Cenke Xu}

\affiliation{Department of Physics, University of California,
Santa Barbara, CA 93106}

\begin{abstract}

We explore the decoherence of the gapless/critical boundary of a topological order, through interactions with the bulk reservoir of ``ancilla anyons."  We take the critical boundary of the $2d$ toric code as an example. The intrinsic nonlocal nature of the anyons demands the strong and weak symmetry condition for the ordinary decoherence problem be extended to the {\it strong or weak gauge invariance conditions}. We demonstrate that in the {\it doubled} Hilbert space, the partition function of the boundary is mapped to two layers of the $2d$ critical Ising model with an inter-layer line defect that depends on the species of the anyons causing the decoherence. The line defects associated with the tunneling of bosonic $e$ and $m$ anyons are relevant, and result in long-range correlations for either the $e$ or $m$ anyon respectively on the boundary in the doubled Hilbert space. In contrast, the defect of the $f$ anyon is marginal and leads to a line of fixed points with varying effective central charges, and power-law correlations having continuously varying scaling dimensions. We also demonstrate that decoherence-analogues of Majorana zero modes are localized at the spatial interface of the relevant $e$ and $m$ anyon decoherence channels, which leads to a universal logarithmic scaling of the R\'enyi entropy of the boundary. 

\end{abstract}


\maketitle

{\bf --- Introduction}

Decoherence of a quantum many-body system describes a loss of information about a quantum state to the environment, and is a generally expected phenomenon in any realistic setting. Instead of thermalization, where a system reaches thermal equilibrium with the environment and all information is lost, 
examples of weak decoherence can be considered in which a pure density matrix $\rho_0$ becomes entangled with a collection of ancilla degrees of freedom, for a \textit{finite} amount of time. The reduced density matrix $\rho^D$ attained by tracing out the ancilla will be a mixed state, which encodes the loss of certainty in $\rho_0$. This type of weak decoherence has a basis in many physical contexts relevant to quantum information 
\cite{de_Groot_2022,altman1,natnishimori,leenishimori,sptdecohere,wfdecohere,bao2023mixedstate,hseihchannel,Entanglement1dCM,KaixiangChern, PurificationTransion2020, BaoChoiEhud2020, jianmeasure, GullansHuseProbes, EntanglementQPTWMSchomerus, HseihDec2}. 
When modeling decoherence, the ancilla degrees of freedom are often treated as local, physical qubits. In two and higher spatial dimensions, however, there exist exotic degrees of freedom in topologically ordered phases that carry point-like energy density but are intrinsically nonlocal, i.e., their creation operator is an extended object. These degrees of freedom are referred to as ``anyons." Indeed, a subsystem coupled to a topological ordered phase may interact with the anyons in its proximate environment and experience new sources of decoherence. Hence, it is the goal of this work to study the consequence of decoherence through a reservoir of ancilla \textit{anyons}. 

As the simplest example of topological order, we will consider the toric code model. The ground-states of the toric code model form robust qubits that can be used to store information
~\cite{kitaev2003}, and decoherence in this model has already attracted considerable interest recently. It is understood that in two-dimensions, topological order cannot survive under thermalization~\cite{ HastingsFiniteT, GerardoFiniteT}. However, this is not generically the case for decoherence resulting from short time exposures to an environment. This process generates decoherence in the form of ``errors" in this state, and the strength of the decoherence can drive a transition past the point at which all errors can be reliably annihilated~\cite{kitaevpreskill,PhysRevX.2.021004}. Interestingly, such transitions can be mapped to critical points in quantities of higher R\'enyi index, e.g. $\tr{(\rho^D)^2}$, and their corresponding statistical mechanics models \cite{bao2023mixedstate, wfdecohere,hseihchannel}.

In this work we explore the decoherence of a critical toric code boundary mediated by anyons of the toric code bulk. In the toric code the three anyons $e,m$ and $f$ have a one-to-one correspondence with certain operators on its critical Ising boundary, namely the spin $\sigma$, the disorder parameter $\mu,$ and the Majorana fermion $\psi,$ respectively. We find that distinct dechoherence channels are generated by the tunneling of each species of anyon between the bulk and the critical boundary. Crucially, on the boundary the anyons are conserved only \textit{weakly}, and this constrains the form of the decohered density matrix $\rho^D$. The effect of the decoherence most dramatically manifests through quantities with a higher R\'enyi index, such as $\cZ_2 = \tr{(\rho^D)^2}$. This quantity can be studied analytically because it can be mapped to two layers of the critical boundary Ising model with an inter-layer line defect that depends on the species of anyon. Finally, we demonstrate that analogs of non-abelian anyons are localized at the spatial interface between the different classes of anyon decoherence.

{\bf --- Preparation: symmetry conditions and the doubled Hilbert space}

We first review some key points related to modeling decoherence. We may prepare an initial pure quantum state of our target system that has a density matrix labelled $\rho_0 = \dyad{\psi}$, where $\ket{\psi}$ is the ground state of a Hamiltonian $H_0$. The decoherence process is essentially a finite-time evolution of the target system and the ancilla degrees of freedom. After the finite-time evolution, the target system will generically be entangled with the ancilla qubits, and tracing out the ancilla qubits will generate a (decohered) mixed state density matrix of the target system $\rho^D$. The decohered mixed state $\rho^D$ may generically be written in terms of a decoherence channel $\varepsilon(\rho_0)$ which maps the pure state of the target system to a mixed state:
\beqn
\rho^D = \varepsilon(\rho_0) = \sum_m \hat{K}_m \rho_0 \hat{K}_m^\dagger \label{channel}.
\eeqn
The operators $\{K_m\}$ are known as Kraus operators and satisfy $\sum_m K_m K_m^\dagger = 1$ for any probability conserving decoherence channel (i.e., those with no post-selection).

For our purposes it is key to distinguish strong and weak symmetry conditions of a density matrix. If the original state $|\psi\rangle$ has a symmetry $G$, namely $G | \psi\rangle = |\psi\rangle$, then the density matrix has a strong (or doubled) symmetry condition: $ \rho_0 = G \rho_0 = \rho_0 G^\dagger$. This strong symmetry condition may be reduced to a weak (or diagonal) symmetry condition after decoherence~\cite{de_Groot_2022,mawang,sptdecohere}, \cx{if the interaction between the target system and the environment enables symmetry charges to tunnel in-and-out of the system while preserving overall charge conservation:} \beqn \rho^D = G \rho^D G^\dagger. \eeqn In our case, the source of decoherence are the ancilla anyons and hence we must consider an additional symmetry distinction: anyons carry charges of a gauge group and thus the strong and weak symmetry conditions are extended to distinctions between strong and weak \textit{gauge invariance}. In particular, if the interaction between the target system (the critical boundary) and the environment (the bulk topological order) enables anyon tunneling in and out of the target system,  tracing out the reservoir of anyons of the environment will yield a mixed density matrix $\rho^D$ that satisfies only a \textit{weak gauge invariance}. As we will see, the concept of weak gauge invariance is crucial for our work and will impose strong constraints on the form of the allowed decoherence channels.

\begin{figure}
    \centering
    \includegraphics[width=0.40\textwidth, height = 2.25in]{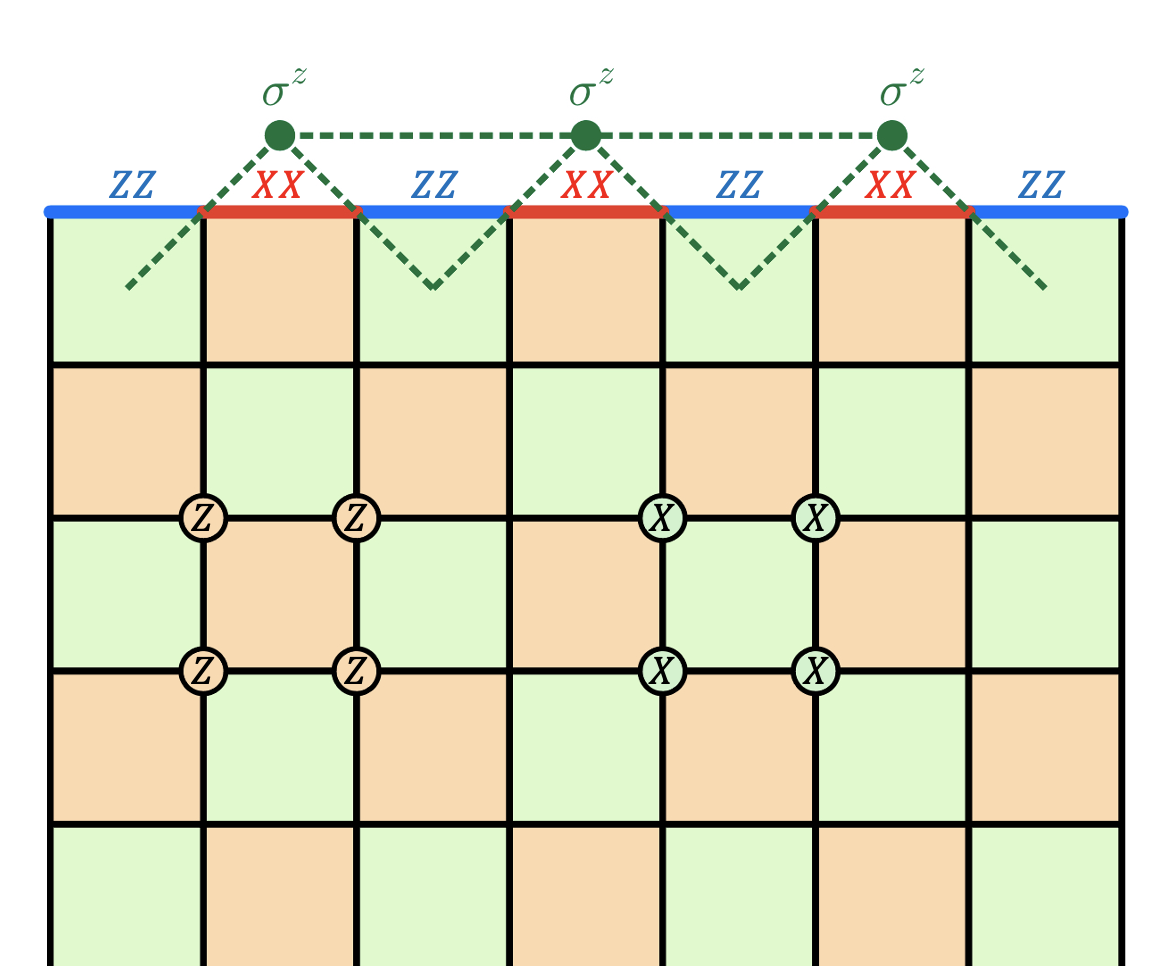}
    \caption{Toric code defined in the bulk, with $A_s$ on plaquettes $s$ (green) and $B_{s'}$ on plaquettes $s'$ (orange). The critical boundary Hamiltonian $H_\text{0} = - \sum_{u \text{ odd} \in \text{bdry}} X_{u}X_{u+\hat{x}} - \sum_{u \text{ even} \in \text{bdry}} Z_u Z_{u+ \hat{x}}$ is shown by the red and blue bars. Upon introducing qubits $(\sigma)$ on the $s$ plaquettes and enforcing $B_{s'} = 1$, the bulk Hamiltonian is dual to a paramagnet $H_\text{tc} = - \sum_s \sigma^x_s$ while the boundary Hamiltonian is dual to $H_0 = - \sum_{v} \sigma^z_{v} \sigma^z_{v+2\hat{x}} - \sum_{v} \sigma^x_{v}$. An extra Zeeman term coupling, $Z$, on the boundary will enable the $e$ anyon to tunnel in-and-out from the bulk and is dual to $- \sum_v \sigma^z_{v} \sigma^z_{v- \hat{y} - \hat{x}} - \sum \sigma^z_{v} \sigma^z_{v - \hat{y} + \hat{x}}$. The index $v$ labels a vertex on the boundary of the dual lattice.}
    \label{tc}
\end{figure}

To reveal the effects of decoherence we consider quantities of higher R\'enyi index. As mentioned above, we will focus on R\'enyi index $n = 2$, e.g., $\cZ_2 = \tr{(\rho^D)^2}$ for the remainder of this work. To compute quantities quadratic in the density matrix, we will make use of the Choi representation~\cite{JAMIOLKOWSKI1972,CHOI1975} which maps a density matrix to a state on a doubled Hilbert space: $\rho^D = \sum_m \hat{K}_m \rho_0 \hat{K}_m^\dagger \rightarrow |\rho^D \rangle \rangle = \hat{\varepsilon}|\rho_0 \rangle \rangle$ where $\hat{\varepsilon} = \sum_m \hat{K}_m \otimes \hat{K}_m^\dagger,$ and $|\rho_0 \rangle \rangle = \ket{\psi} \otimes \ket{\psi}$. We note that the mapping takes a mixed state density matrix $\rho^D$ to a {\it pure state wave function} in the doubled Hilbert space which is given by the operator $\hat{\varepsilon}$ acting on two copies of $\ket{\psi}$~\cite{sptdecohere,bao2023mixedstate}. Importantly, the state $|\rho^D \rangle \rangle$ has norm $\langle \langle \rho^D | \rho^D \rangle \rangle = \tr{(\rho^D)^2},$ and expectation values of local operators $\hat{\mathcal{O}}$ computed in $\cZ_2$ become quantum expectation values of the state $|\rho^D \rangle \rangle$: 
\beqn
\tr{ \hat{\mathcal{O}}\rho^D \hat{\mathcal{O}}' \rho^D} = \langle \langle \rho^D|\hat{\mathcal{O}} \otimes \hat{\mathcal{O}}'|\rho^D \rangle \rangle. 
\eeqn

{\bf --- Decoherence through the $e$ and $m$ anyons}

Our target system is the critical boundary of the toric code topological order. We will first construct an explicit lattice model for the decoherence of the critical boundary generated by tunneling of the $e$ anyons. Consider the square lattice having qubits on its vertices and its (smooth) boundary. The bulk Hamiltonian for the toric code is
\beqn
    H_{\text{tc}} = - \sum_{s}  A_s - \sum_{s'} B_{s'} \label{tcH}. 
\eeqn
Here $A_s = \prod_{u \in s}X_u$ and $B_{s'} = \prod_{u \in {s'}}Z_u$ with $s$,$s'$ labeling the two different types of checkerboard plaquettes of the square lattice and $u$ labeling the vertices as shown in Fig.~\ref{tc}. In this representation, the $e$ anyons and $m$ anyons both live on plaquettes and correspond to $A_s = -1$ and $B_{s'} = -1$ excitations respectively. Our target system will be a critical Ising model in $1d$ (one spatial dimension) on the boundary. The Hamiltonian of this model (also shown in Fig.~\ref{tc}) is
\beqn
 H_\text{0} = - \sum_{u \text{ odd} \in \text{bdry}} X_{u}X_{u+\hat{x}} - \sum_{u \text{ even} \in \text{bdry}} Z_u Z_{u+ \hat{x}} \label{boundaryH}.
\eeqn
The boundary $X_u X_{u+\hat{x}}$ and $Z_u Z_{u+ \hat{x}}$ terms are the boundary $e$ and $m$ anyon densities respectively.

We aim to turn on interactions which tunnel $e$ anyons between the bulk and the boundary. To accomplish this we introduce a Zeeman field $H_\text{Zee}=- \sum_{u \in \text{bdry}}Z_u$ along the boundary. This term anticommutes with both the $e$ anyon terms in the bulk ($A_s$) and the boundary ($X_u X_{u+\hat{x}}$). In other words, this Zeeman field enables $e$ anyon tunneling between the boundary and the bulk, while the combined bulk-boundary system still conserves the number of $e$ anyons. After tracing out the bulk, the $e$ anyons are conserved only weakly on the boundary which constrains the form of $\rho^D$. 

To elucidate the expected form of the decoherence on the boundary, it is useful to represent the boundary Hamiltonian in a more standard form. The local Gauss' law $B_{s'} = 1, \forall s'$, constrains the density of $m$ anyons to vanish and can be enforced after turning on the Zeeman field on the boundary.  In the fashion of the standard Ising duality in $2d$, every $s$-type plaquette (vertex of the dual lattice) now supports qubits denoted by $\sigma$, for which $Z_u$ is represented as the product of its two neighboring $\sigma^z$ operators. That is, $Z_{u \in \text{bulk}} = \sigma^z_{s_1} \sigma^z_{s_2}$ with $(s_1,s_2)$ being nearest-neighbors on the dual lattice on link $u$, and likewise for $Z_{u \in \text{bdry}} = \sigma^z_{s} \sigma^z_{v}$ but with $v$ specifically labelling a vertex on \textit{the boundary of the dual lattice}. As illustrated in Fig.~\ref{tc}, Eq.~\ref{boundaryH} is mapped to $H_0= -\sum_{v} \sigma^z_{v} \sigma^z_{v+2 \hat{x}} - \sum_{v} \sigma^x_{v}$ while the bulk toric code Eq.~\ref{tcH} is mapped to a trivial paramagnet, i.e. $H_{tc}=- \sum_{s} \sigma^x_{s}$. The Zeeman field enters as an Ising interaction that couples the critical boundary and the paramagnetic bulk, $H_\text{Zee}= - \sum_v \sigma^z_{v} \sigma^z_{v - \hat{y} - \hat{x}} - \sum \sigma^z_{v} \sigma^z_{v - \hat{y} + \hat{x}}.$ After this mapping the expectation value of an $e$ anyon Wilson line having endpoints on the boundary reduces to the two-point function of $\sigma^z_{v}$ operators at the endpoints.  

Upon tracing out the degrees of freedom other than the qubits $\sigma^x$ on the boundary, $\rho^D$ takes the form of a decoherence channel in Eq.~\ref{channel} acting on $\rho_0 = \dyad{\psi}$ (the pure, ground state density matrix of the critical Ising model~\footnote{Without the Zeeman field, the boundary qubits $\sigma^x$ are decoupled from the bulk, hence the density matrix of the boundary $\sigma^x$ after integrating out the the bulk would still be a pure $(1+1)d$ CFT density matrix. Without the Zeeman field the product of $\sigma^x$ along the boundary must be 1, which is an extra global constraint on the boundary.}). The form of the Zeeman field interaction constrains that the local decoherence channel at each boundary site $v$ is well-modeled by:
\beqn
\varepsilon_v(\rho_0) = (1-p) \rho_\text{0} + p\sigma^z_v \rho_\text{0} \sigma^z_v \label{eDecChannel},
\eeqn where $p\in [0,1]$ is a probability.
The full decohered density matrix is given by the composition of the local decoherence channels: $\rho^D \sim \otimes_v \varepsilon_v(\rho_0)$. Intuitively, like the anyons, $\sigma^z_{v}$ is not a gauge invariant object, and $\sigma_v^z \ra - \sigma_v^z $ is part of a gauge redundancy. Since we enabled $e$ anyon tunneling between the boundary and the bulk, the boundary should have only a weak gauge invariance, hence Eq.~\ref{eDecChannel} is only invariant under $\sigma^z \ra - \sigma^z $ on both sides simultaneously.

In the doubled Hilbert space formalism, the density matrix $\rho^D$ is mapped to a state $|\rho^D \rangle \rangle$ in the doubled Hilbert space. For the decoherence channel in Eq.~\ref{eDecChannel} acting on $\rho_0$ (the density matrix of the critical Ising ground state), the decohered density matrix state is
\begin{equation*}
|\rho^D \rangle \rangle \sim \prod_v\left( 1 + \frac{p}{1 - p} \sigma^z_v \otimes \sigma^z_v \right) |\rho_0 \rangle \rangle \sim e^{g \sum_v \sigma^z_v \otimes \sigma^z_v} |\rho_0 \rangle \rangle \label{eDecoherencestate}
\end{equation*}
where $g = \text{arctanh}(\frac{p}{1-p})$. The decoherence operator $\hat{\varepsilon} = e^{g \sum_v \sigma^z_v \otimes \sigma^z_v}$ represents the coupling between the two copies of the critical Ising boundary. We can now map the partition function for R\'enyi index $n = 2$, $\cZ_2 = \tr{(\rho^D)^2} = \tr{e^{2g \sum_v \sigma^z_v \otimes \sigma^z_v} | \rho_0 \rangle \rangle \langle \langle \rho_0 |}$ to a $2d$ statistical mechanics model, i.e., the partition function of two copies of the Ising model at the critical temperature and an inter-layer defect of coupling strength $g' \sim g$ on a line. The classical Hamiltonian of this model is
\beqn
 \beta \cH &=& \beta \cH_0 - g' \sum_{\tau = 0} \sigma_{i} \tilde{\sigma}_i, \cr\cr
\text{ with }\beta \cH_0 &=& -J_c \sum_{\langle i,j \rangle} \sigma_i \sigma_j  -J_c \sum_{\langle i,j \rangle} \tilde{\sigma}_i \tilde{\sigma}_j, \label{eDecohereEuc}
\eeqn
\begin{center}
\begin{figure}
    \centering
    \includegraphics[width=0.47\textwidth]{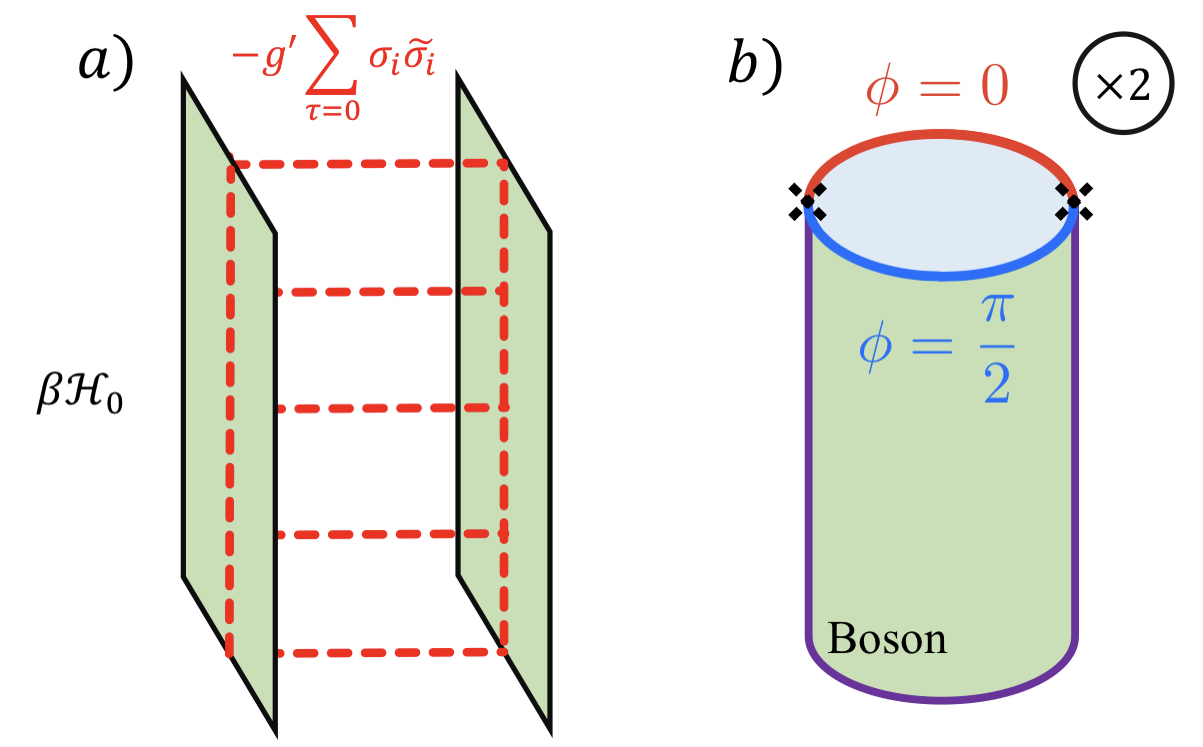}
    \caption{$a)$ Two copies of the critical Ising model, with an inter-layer defect along the red line that couples the spins of each layer together of $-g' \sum_{\tau = 0} \sigma_i \tilde{\sigma}_i$ (Eq.~\ref{eDecohereEuc}). $b)$ In the doubled Hilbert space, $\cZ_2 = \tr{(\rho^D_{em})^2}$ for the $em$-kink is equivalent to two cylinder partition functions of a compact boson with a $\frac{\pi}{2}$ domain wall on the boundary.}
    \label{Defect/bosonDW}
\end{figure}
\end{center}
where the index $i = (x,\tau)$ labels the vertex of the $2d$ square lattice. The direction $\tau$ can be interpreted as Euclidean time. The defect is on the line $\tau=0$, and the spins $\sigma$ and $\tilde{\sigma}$ are the spins on each of the two layers (which are mapped from $\sigma^z_v \otimes 1$ and $1 \otimes \sigma^z_v$ in the quantum model respectively). This line defect represented by the term $-g' \sum_{\tau = 0} \sigma_i \tilde{\sigma}_i$ explicitly breaks the doubled $Z_2$ symmetries of the  two layers down to the diagonal subgroup $Z_2^\text{diag} : \sigma,\tilde{\sigma} \rightarrow -\sigma,-\tilde{\sigma}$.

For two layers of the Ising CFT, the dimension of the $\sigma \tilde{\sigma}$ operator is $\Delta_{\sigma \tilde{\sigma}} = \frac{1}{4}$, hence $g'$ is strongly relevant. Under bosonization, two copies of Ising CFT are mapped to a single compact boson $\phi\sim \phi+2\pi,$ with the identification $- g'\int \dd{x}  \sigma(x,0) \tilde{\sigma}(x,0) 
\sim - g'\int \dd{x}   \cos(\phi(x,0))$ along the defect. A relevant $g'$ will pin $\phi = 0$ at the defect, and effectively cut the system into two halves. As shown in Ref.~\cite{MasakiDefect}, for each side of the system, $\phi = 0$ corresponds to the ferromagnetic boundary condition which exhibits long range-order of $\sigma^z$ \footnote{Since $\sigma^z \ra - \sigma^z$ is a gauge redundancy in our case, the two FM boundary conditions are physically identical.}. The long-range order of the spins $\sigma^z$ (i.e., in the doubled Hilbert space $\tr\{ \sigma^z_v \rho^D \sigma^z_{v'} \rho^D \},\tr\{ \sigma^z_v  \sigma^z_{v'} (\rho^D)^2 \} \ra \text{const}$) has the interpretation as long-range order of the $e$ anyon Wilson line correlation functions. One can also verify that, under the decoherence from tunneling $e$ anyons, the $m$ anyon correlation function becomes short-ranged. In a sense, the effects of decoherence in the doubled Hilbert space are ``strong" as $g'$ (which depends on $p$ in Eq.~\ref{eDecChannel}) flows to infinity. 

The construction of the lattice model for decoherence from  $m$ anyon tunneling between bulk and boundary is similar to what we have discussed for the $e$ anyon case. Instead of the Zeeman field, a transverse field $H_\text{TF}=- \sum_u X_u$ is turned on along the boundary to induce $m$ anyon tunneling. Indeed, this term violates both bulk and boundary $m$ anyon conservation separately, while the $m$ anyon remains conserved in the entire system. After tracing out the bulk, the resulting decoherence channel will be the Kramers-Wannier dual of Eq.~\ref{eDecChannel}. Consequently, this new decoherence channel drives long-range order of the (doubled pair of) disorder parameters $\mu(x,\tau)$ and $\tilde{\mu}(x,\tau)$, i.e., long-range correlation of the $m$ anyons, but short-range correlation of the $e$ anyons.

{\bf --- Decoherence through the $f$ anyon}

The $f$ fermion in toric code is the fusion of an $e$ anyon and an $m$ anyon, and carries the gauge charge of both anyons. Analogously, in the critical Ising model, the fusion of a spin and a domain wall is a Majorana fermion, which is essentially inherited from the $f$ fermion of the toric code. Without making direct reference to a lattice model, the resulting local decoherence channel is constrained by weak gauge invariance conditions to be of the form
\beqn
\epsilon_v(\rho_0) = (1-p) \rho_0+p \sigma^z_v \mu^z_v \rho_0 \sigma^z_v \mu^z_v, \label{fDecchannel}
\eeqn
where $\rho^D \sim \otimes_v \varepsilon_v(\rho_0)$. The density matrix state in the doubled Hilbert space is 
\beqn
|\rho^D \rangle \rangle \sim e^{g \sum_v \sigma^z_v \mu^z_v \otimes \sigma^z_v \mu^z_v} |\rho_0 \rangle \rangle.
\eeqn
The resulting decoherence can be most conveniently studied by bosonizing two copies of the critical Ising CFT to one compact boson, as mentioned above, under the following prescription
\beqn
\sigma(x,\tau) \tilde{\sigma}(x,\tau) &\rightarrow& \cos(\phi(x,\tau)), \cr
\mu(x,\tau) \tilde{\mu}(x,\tau) &\rightarrow& \sin(\phi(x,\tau)), \cr
\mathcal{E}(x,\tau) + \tilde{\mathcal{E}}(x,\tau) &\rightarrow& \cos(2\phi(x,\tau)) \label{bosonization},
\eeqn
where $\mathcal{E}$ and $\tilde{\mathcal{E}}$ are the energy density operators which tune the sytsem away from Ising criticality. The partition function of two copies of the Ising CFT becomes the Euclidean time path integral for the boson $\cZ_2 \sim \int \mathcal{D} \phi \text{ } e^{-S_0}, S_0 = \frac{1}{2 \pi} \int \dd[2]{x} (\partial\phi)^2$. Decoherence  enters as an inter-layer line defect $\sim -g'\int \dd{x} \sigma(x,0) \mu(x,0) \tilde{\sigma}(x,0) \tilde{\mu}(x,0)$, and is mapped to the bosonized defect 
\beqn
S_\text{defect} = -g' \int \dd{x} \sin(2\phi(x,0)). \label{fDecDefect}
\eeqn
The total action is $S = S_0+S_\text{defect}, $ and the Kramers-Wannier duality of both Ising CFTs corresponds to the $\phi \rightarrow \frac{\pi}{2} - \phi.$ We note that $S_\text{defect}$ is invariant under the doubled duality transformation, as it should be. Also, since $\cos(\phi)$ and $\sin(\phi)$ both have scaling dimension $\Delta = \frac{1}{4}$ at Ising criticality, $\sin(2\phi)$ has scaling dimension $\Delta = 1$, making $S_\text{defect}$ exactly marginal. 

To simplify $S_\text{defect}$ we perform the rotation $\phi \ra \phi + \pi/4$, which makes $S_\text{defect} \sim -g'\int \dd{x} \cos(2\phi) $. This operator is simply identified as $\mathcal{E}(x) + \tilde{\mathcal{E}}(x) $ from Eq.~\ref{bosonization}. Helpfully, the problem of a defect line of $\cos(2\phi(x))$ at $\tau = 0$ corresponds to another problem that has recently been studied: what happens to a $1d$ Ising critical wave function when one weakly measures the transverse field \cite{hseihchannel,Entanglement1dCM}. This marginal defect leads to two consequences: \textit{(1)} it gives the system an effective central charge, which we elaborated upon in the SM,
and \textit{(2)} it changes the scaling dimension of $\sigma(x,\tau) \tilde{\sigma}(x,\tau)$ and $\mu(x,\tau) \tilde{\mu}(x,\tau)$ to \beqn
\Delta_{\sigma \tilde{\sigma}} = \Delta_{\mu \tilde{\mu}} = \frac{4}{\pi^2} \text{arctan}^2 \left(e^{-2|g'|}\right) \label{FermionDim}.
\eeqn
While the two-point functions of the bound state operators $\sigma(x,\tau) \tilde{\sigma}(x,\tau)$ and $\mu(x,\tau) \tilde{\mu}(x,\tau)$ remain power-law correlated, the spin and disorder parameter two-point functions, in the same layer, individually are rendered short-range (please refer to the SM).

{\bf --- The interface of $e$- and $m$-decoherence}

The boundary of topological order can host exotic zero dimensional defects. For example, the domain-wall between an $e$ and $m$ condensed region of the boundary of toric code can host a Majorana zero mode, which is non-abelian, even though all bulk anyons are abelian. This phenomenon also has an analog in our decoherence problem. We can consider a scenario of decoherence through the $e$ anyons on half the critical boundary, and decoherence via $m$ anyons on the other half; a geometry that we refer to as the $em$-kink. To illustrate, we place the toric code and its critical boundary on a \textit{large} semi-infinite cylinder of circumference $L$ (i.e., $L \rightarrow \infty$). In this cylindrical geometry, there is an $em$-kink at the origin of the boundary and another one a distance $L/2$ away on the boundary. As we will demonstrate, this $em$-kink endows the second R\'enyi entropy with a universal logarithmic scaling, where the coefficient is proportional to twice the scaling dimension of an operator which is analogous to a pair of Majorana zero modes.

The decoherence channel corresponding to the $em$-kink takes the following form:
\beqn
\varepsilon_v(\rho_0) \sim 
\begin{cases}
   (1-p)\rho_0 + p \sigma^z_v \rho_0 \sigma^z_v & L/2 > v >0 \\
   (1-p) \rho_0 + p \mu^z_v \rho_0 \mu^z_v & -L/2 < v < 0 .
\end{cases}
\label{emDec}
\eeqn
This is an exotic decoherence channel which is achieved by leveraging that ancilla anyons can entangle with mutually non-local degrees of freedom of the boundary, i.e., both $e$ and $m$ anyons can tunnel in their respective regions. We aim to compute the 2nd Renyi entropy of the decohered boundary with the two $em$-kinks,
\beqn
\cF_{em-\text{kink}} = - \log \bigg ( \frac{\tr{(\rho_{em}^D)^2}}{\tr{(\rho_{e}^D)^2}} \bigg ) \label{Freeenergy}.
\eeqn
Here, $\rho^D_{em} \sim \otimes_v \varepsilon_v(\rho_0)$ is the density matrix under decoherence in Eq.~\ref{emDec}, while $\rho^D_e$ is generated by dechoherence in  Eq.~\ref{eDecChannel}. Using bosonization, $\tr{(\rho^D_{em})^2}$  is mapped to the partition function of two layers of the Ising CFT on an infinite cylinder of circumference $L$ in the spatial direction, with an inter-layer defect. The inter-layer defect is bosonized as
\beqn
S_\text{defect} &=& -g' \int \limits_{-L/2}^0 \dd{x} \sin(\phi(x,0)) \cr \cr&-& g' \int \limits_{0}^{L/2} \dd{x} \cos(\phi(x,0)). \label{emDefect}
\eeqn
This defect is relevant and pins the compact boson to $\phi = \frac{\pi}{2}$ for $L/2 > x > 0,$ and pins $\phi = 0$ for $-L/2 < x < 0$. This effectively cuts the connection of the compact boson between the two sides of the defect, and hence results in \textit{two} compact bosons on the semi-infinite cylinder. Furthermore, on each cap of the semi-infinite cylinder there is a $\frac{\pi}{2}$ domain wall between the $\phi = 0$ and the $\phi = \frac{\pi}{2}$ boundary conditions, and another opposite domain wall at a distance $L/2$ away at the other kink, as shown in Fig.~\ref{Defect/bosonDW}b. Hence an $em$-kink in the decoherence problem is equivalent to a $\frac{\pi}{2}$ domain wall on the boundary of the two compact bosons with boundary condition $\phi = 0$.

Hence, we have reduced or problem to studying the class of boundary operators called boundary condition changing operators (BCCOs) for a given conformal boundary condition of a CFT. For the compact boson CFT, there are primary BCCOs $\hat{\mathcal{B}}_x(\alpha,\beta)$ at position $x$ on the boundary that take a generic Dirichlet boundary condition $\alpha$ to another one, $\beta$. For the purposes of our discussion, they may be regarded as the creation or annihilation operators of domain walls on the boundary of the cylinder. 
The quantity $\tr\{(\rho^D_{em})^2\}/\tr{(\rho^D_e)^2}$ is mapped to the square of the correlation function of BCCOs $\langle \hat{\mathcal{B}}_{x = 0}(0, \frac{\pi}{2}) \hat{\mathcal{B}}_{x = L/2}(\frac{\pi}{2},0) \rangle$. This constrains the free energy Eq.~\ref{Freeenergy} to take the following form 
\beqn
\cF_{em-\text{kink}} &=& - \log \bigg \langle \hat{\mathcal{B}}_{x=0} \Big (0, \frac{\pi}{2} \Big)  \hat{\mathcal{B}}_{x=L/2} \Big (\frac{\pi}{2},0 \Big) \bigg \rangle^2, \cr\cr
&\sim& - 4  \Delta_{\hat{\mathcal{B}}} \log (\frac{L}{2}).
\eeqn
In this expression, $\Delta_{\hat{\mathcal{B}}}$ is the dimension of the BCCOs $ \hat{\mathcal{B}}_x(0, \frac{\pi}{2})$. The dimension of a generic BCCO $\hat{\mathcal{B}}_x(\alpha,\beta)$ has been computed in Ref.~\cite{MasakiDefect} for $\Delta \phi$ the smallest distance between $\beta$ and $\alpha$ on the unit-circle, as
\beqn
\Delta_{\hat{\mathcal{B}}} = \frac{1}{2} \bigg ( \frac{ \Delta \phi}{ \pi} \bigg )^2. \label{BCCOdim}
\eeqn 
For the case here, we have $\Delta_{\hat{\mathcal{B}}} = \frac{1}{8}$, such that $\cF_{em-\text{ kink}} \sim - \frac{1}{2} \log ( \frac{L}{2})$. Hence, the operators $\hat{\mathcal{B}}_x(0, \frac{\pi}{2}),\hat{\mathcal{B}}_x(\frac{\pi}{2}, 0)$ are analogous to a pair of distinct Majorana zero modes. We present additional discussion of the calculation in the SM, which may also be found in the appendix of Ref.~\cite{MasakiDefect}.

{\bf --- Summary and discussion}

In this work, we discussed the mixed state density matrix of the critical boundary of a toric code model after allowing decoherence caused by different types of anyons tunneling into/from the bulk. The anyons play the same role as the reservoir of ancilla qubits in conventional decoherence approaches. The mixed state density matrix is mapped to a pure quantum state in the doubled Hilbert space, and anyon decoherence generates an interlayer line defect of two copies of the $2d$ classical Ising CFT. The type of the line defect depends on the anyon type that causes the decoherence, and can lead to different results. 

We were able to use techniques and results of from previous studies on the boundary of the Ising CFT to calculate the second Re\'nyi entropy. Futhermore, we find that the physics here is far richer than decoherence on a physical quantum Ising chain generated by entangling it with ordinary ancilla qubits. For example, the ordinary ancilla qubits can entangle with the Ising spin $\sigma^z$ or the transverse field $\sigma^x$ of the quantum Ising chain, but not with the disorder parameter, or the $\psi$ Majorana fermion, as these are nonlocal objects of the Ising chain. Hence, anyon decoherence opens the gate to a much broader playground of systems that need only obey weak gauge invariance when decohering. 

Our study can be naturally extended to the anyon-decohered gapless boundary of other topological orders. In recent years, the symmetry of a boundary of topological order has been systematically shown to inherit from the conservation and fusion rules of anyons in the topological bulk. These symmetries are generally referred to as ``categorical symmetry"~\cite{cat1,cat2} and we expect that the strong and weak \emph{categorical} symmetry conditions could serve as a general guiding principle for future studies. 

{\bf --- Acknowledgements}

C.X. acknowledge the supports from the Simons Foundation through the Simons investigator program. T.L.H. acknowledges support from the U.S. Office
of Naval Research (ONR) Multidisciplinary University Research Initiative (MURI) under Grant No. N00014-20-1-2325 on Robust Photonic Materials with Higher-Order Topological
Protection. The authors would like to extend their gratitude to Tim Hsieh, Chao-Ming Jian, Max Metlitski and Masaki Oshikawa for inspiring discussions. N.M.J. would also like to thank Zhehao Zhang.

\bibliography{main}

\newpage
\onecolumngrid
\appendix

\section{Appendix A: Weak measurement; scaling dimensions and effective central charge for $f$ anyon tunneling in Toric Code}

It is useful to review the decoherence channel associated with direct weak measurement of the transverse field in the critical Ising model. This problem has been studied in the context of decoherence physics before (see Ref.~\cite{hseihchannel,Entanglement1dCM}). In our problem, this type of decoherence involves directly weakly measuring the gauge-invariant charge $\sigma^x_v$. We can also allow for the option of $\textit{post-selection}$ of the charge on the defect line directly with projectors $\mathcal{P}_\pm = \big ( \frac{1\pm\sigma^x_v }{2} \big )$
\beqn
    \varepsilon_v(\rho_0) = (1-p_+-p_-) \rho_0  + p_+\mathcal{P}_+\rho_0 \mathcal{P}_+ + p_- \mathcal{P}_- \rho \mathcal{P}_-.
\eeqn
we have allowed for \textit{post-selection} of the measurement result of $\sigma^x$. This decoherence channel has a nice interpretation as the following random measurement protocol performed on every vertex, where we take one of three potential actions: $\textit{1)}$ we make no measurement with probability $1 - p_+ - p_-$, $\textit{2)}$ we measure a charge $\sigma^x_v = 1$ with probability $p_+$, and $\textit{3)}$ we measure no charge $\sigma^x_v = -1$ with probability $p_-$. In the case that $p_+ = p_-$, corresponds to no post-selection at all as there is an equal probability to find either $\sigma^x_v = 1$ or $\sigma^x_v = -1$. Note that we could not have measured the anyons on the boundary directly, as they both carry gauge charge. The decohered state in the doubled Hilbert space and the corresponding $2d$ statistical mechanics model is
\beqn
    | \rho^D \rangle \rangle &\sim& e^{\frac{(g_+-g_-)}{2} \sum_v 1 \otimes \sigma^x_v + \sigma^x_v \otimes 1 + \frac{g_+ + g_-}{2} \sum_v \sigma^x_v \otimes \sigma^x_v }  |\rho_0 \rangle \rangle \\
   \Rightarrow \beta \cH = \beta \cH_0 &-& g' \sum_{x} \sigma_{x, 0} \sigma_{v,1} -g'\sum_x \tilde{\sigma}_{x,0} \tilde{\sigma}_{x,1} -g''\sum_x \sigma_{x,0} \sigma_{x,1} \tilde{\sigma}_{x, 0} \tilde{\sigma}_{x,1} \label{MeasureEucH}.
\eeqn
with $g_\pm = \log \big ( \frac{1 - p_\mp}{1- p_+ - p_-} \big )$, $g' \sim g_+ - g_-$, $g'' \sim g_+ + g_-$ and $\mathcal{H}_0$ given in the main text. It is useful to remind ourselves that the vertices of this $2d$ model are labeled by $(x,\tau)$.

The first two defects with coupling $g'$ are $\textit{intra-layer}$ defects. They correspond to a ladder of links of the Ising model, on which the nearest-neighbor Ising interaction is decreased from $-J_c$ to $-J_c - g'$. In the Ising CFT language, this corresponds to a line defect of the energy density operator $\sim -g' \int \dd{x}\left( \mathcal{E}(x,0)+\tilde{\mathcal{E}}(x,0)\right)$, which is exactly marginal. The second defect with coupling $g''$ corresponds to a $\textit{inter-layer}$ defect which couples these two lines together. At best, this inter-layer defect can only renormalize the coupling $g'$ in the continuum; any other possible terms that arise from it will be irrelevant. Consequently, the continuum version of Eq.~\ref{MeasureEucH} is equivalent to two copies of the Ising CFT with an energy density defect line. In bosonization, the energy line defect is
\beqn
S_\text{defect} = -g' \int \dd{x} \cos(2\phi(x,0)) \label{energydefectbos}.
\eeqn
The energy density line defect in the Ising CFT is an old problem (for example, see Refs.~\cite{McCoyEnergyDefect,BarievEnergyDefect, BrownEnergyDefect} among many others). The conformal boundary conditions for the folded theory, the dimensions of all primary operators, defect entropies and the effective entanglement entropy are all known \cite{MasakiDefect,Entanglement1dCM}. In particular, the dimension of the spins and disorder parameters in each layer is
\beqn
\Delta_\sigma = \Delta_{\tilde{\sigma}} &=& \frac{2}{\pi^2} \text{arctan}^2 (e^{-2g'}),  \\ \label{spindimWM}
\Delta_\mu = \Delta_{\tilde{\mu}} &=& \frac{2}{\pi^2} \text{arctan}^2 (e^{2g'}). \label{disorderdimWM}
\eeqn
The two layers are completely decoupled in this problem. In bosonization, $\sigma(x,0) \tilde{\sigma}(x,0) \rightarrow \cos(\phi(x,0))$ and $\mu(x,0) \tilde{\mu}(x,0) \rightarrow \sin(\phi(x,0))$. As such, the dimension of these vertex operators in the bosonic theory are just given by twice the above equations, respectively. Lastly, the entanglement entropy retains its critical scaling, $S_\text{EE}(\ell) \sim \frac{c_\text{eff}}{3} \log \ell$ ($\ell$ is the length of the chosen subsystem), but now with an effective central charge that decreases with the measurement strength
\beqn
c_\text{eff}(g') = - \frac{6}{\pi^2}\{[(1+s)\log(1+s)+(1-s)\log(1-s)]\log(s) + (1+s)\text{Li}_2(-s)+(1-s)\text{Li}_2(s)\} \label{ceff}
\eeqn
for $s = 1/(\text{cosh}(4g'))$. This value is twice the result in Ref.~\cite{Entanglement1dCM} as we have two copies of the energy density line defect problem.

As explained in the main text, the effect of fermion decoherence in the doubled Hilbert space problem is the introduction of the line defect $S_\text{defect} = -g' \int \dd{x} \sin(2\phi(x,0))$. This is related to Eq.~\ref{energydefectbos} by the $\U(1)$ transformation $\phi \rightarrow \phi+\frac{\pi}{4}$. This takes this $\langle \sigma(x,0) \tilde{\sigma}(x,0) \sigma(x',0) \tilde{\sigma}(x',0) \rangle = \langle \cos(\phi(x,0)) \cos(\phi(x',0)) \rangle$ correlation to the following
\beqn
\langle \cos(\phi(x,0)) \cos(\phi(x',0)) \rangle &\longrightarrow& \sim \langle [\cos(\phi(x,0)) - \sin(\phi(x,0))][\cos(\phi(x',0)) - \sin(\phi(x',0))] \rangle,\cr\cr
&=& \langle \cos(\phi(x,0)) \cos(\phi(x',0)) \rangle + \langle \sin(\phi(x,0)) \sin(\phi(x',0)) \rangle.
\eeqn
The cross terms necessarily vanish as they are odd under reflection symmetry of the boson, $\phi \rightarrow - \phi$ which Eq.~\ref{energydefectbos} preserves. Hence the scaling dimension of $\Delta_{\sigma \tilde{\sigma}}$, is

\beqn
\Delta_{\sigma \tilde{\sigma}} = \min \bigg \{ \frac{4}{\pi^2} \arctan^2(e^{-2g'} ), \frac{4}{\pi^2} \arctan^2(e^{2g'} )  \bigg \} = \frac{4}{\pi^2} \arctan^2(e^{-2|g'|} ),
\eeqn
which is Eq.~\ref{FermionDim}. A similar calculation shows that $\Delta_{\mu \tilde{\mu}}$ is also given by Eq.~\ref{FermionDim}. Lastly, because the $S_\text{defect} = -g' \int \dd{x} \sin(2\phi(x,0))$ is related to Eq.~\ref{energydefectbos} by a $\U(1)$ transformation, they have the same partition function: the entanglement entropy of each theory must be the same and hence the same effective central charge given in Eq.~\ref{ceff}.

\section{Appendix B: Calculation of short-range correlation functions for $f$ anyon tunneling in Toric Code}

Here, we will compute spin and disorder parameter correlation functions (e.g. $\langle \sigma_v \sigma_{v'} \rangle$) under $f$ fermion decoherence, and show that they are short-range. Recall that the pertinant decoherence channel is $\varepsilon_v(\rho_0) = (1-p)\rho_0 + p \sigma^z_v \mu^z_v \rho_0 \sigma^z_v \mu^z_v$.  The disorder parameter $\mu^z_v = \prod_{u<v} \sigma^x_u$ creates, or annihilates, a domain wall. The decohered density matrix $\rho^D \sim \otimes_v \varepsilon_v(\rho_0)$ in the doubled Hilbert space formalism is mapped to the state

\beqn
    |\rho^D \rangle \rangle \sim e^{g \sum_v \sigma^z_v \mu_v^z  \otimes \tilde{\sigma}_v^z \tilde{\mu}^z_v } | \rho_0 \rangle \rangle \label{fdecstateapp}.
\eeqn
The state $|\rho_0 \rangle \rangle = \ket{\psi} \otimes \ket{\psi}$, is two copies of the critical Ising model ground-state. We will keep the tensor product implicit from here on out to simplify notation. The critical Ising model can be mapped to a model of free Majorana fermions via the Jordan-Wigner transformation. As stated in the main text, the required mapping between spin operators and fermion operator for two copies of the critical Ising model are as follows:
\beqn
    \sigma^z_v &\sim &e^{i \pi \sum_{u < v} c_u^\dagger c_u} (c_v^\dagger +c_v) \text{ and } \mu^z_v \sim e^{i \pi \sum_{u < v} c_u^\dagger c_u} \cr\cr
    \tilde{\sigma}^z_v &\sim& \kappa e^{i \pi \sum_{u < v} a_u^\dagger a_u} (a_v^\dagger +a_v) \text{ and }
    \tilde{\mu}^z_v \sim e^{i \pi \sum_{u < v} a_u^\dagger a_u}.
\eeqn
The fermions $a_v,c_v$ for each Ising model obey the standard anticommutation relations. The Klein factor $\kappa$ appears in order to enforce that the fermionic representations of the spins and disorder parameters obey the correct commutation relations. It is nothing but the $Z_2$ Ising charge of one of the critical Ising models, $\kappa = \prod_v \sigma^x_v = e^{i \pi \sum_v c_v^\dagger c_v}$. Note that $c_v^\dagger + c_v$ and $e^{i \pi c_v^\dagger c_v}$ anticommute. This may be checked by using the fact that $e^{i \pi c_v^\dagger c_v} = 1+(e^{i \pi} - 1)c_v^\dagger c_v = 1 - 2c_v^\dagger c_v = (c_v^\dagger+c_v)(c_v^\dagger - c_v)$, and that $(c_v^\dagger+c_v)$ and $(c_v^\dagger - c_v)$ also mutually anticommute. The Klein factor is important because it enforces that the representation of spin operators in terms of fermions have the correct commutation relations. The state $|\rho^D \rangle \rangle$ becomes in terms of fermions 
\beqn
    |\rho^D \rangle \rangle \sim e^{g \sum_v  (c_v^\dagger + c_v) (a_v^\dagger + a_v) \kappa} |\rho_0 \rangle \rangle.
\eeqn
The operator $\hat{\varepsilon} = e^{g \sum_v (c_v^\dagger+c_v)(a_v^\dagger+a_v) \kappa}$ is Hermitian (as required) due to the inclusion of the Klein factor as well. We aim to compute $\langle \sigma^z_v \sigma^z_{v'} \rangle = \frac{\langle \langle \rho^D | \sigma^z_v \sigma^z_{v'} |\rho^D \rangle \rangle}{\langle \langle \rho^D | \rho^D \rangle \rangle}$. Note that the spin operator does not commute with a number of terms on the order of the system size. As a result, we cannot reliably compute $\langle \sigma^z_v \sigma^z_{v'} \rangle$ as the spin-spin correlation function in $\cZ_2 = \tr{(\rho^D)^2}$ by commuting it with $\hat{\varepsilon},\hat{\varepsilon}^\dagger$. To compute this correlation function, we must genuinely evaluate the expectation value in $|\rho^D \rangle \rangle$. In what follows, we take $v' \geq v$ without loss of generality. The numerator can be written in terms of fermions as

\begin{align*}
    \langle \langle \rho^D|\sigma^z_v \sigma^z_{v'}|\rho^D \rangle \rangle =  \langle \langle \rho_0 |e^{g \sum \limits_u  (c_u^\dagger + c_u) (a_u^\dagger + a_u) \kappa} (c_v^\dagger +c_v) e^{i \pi \sum \limits_{v \leq w < v'} c_w^\dagger c_w} (c_{v'}^\dagger +c_{v'}) e^{g \sum \limits_u  (c_u^\dagger + c_u) (a_u^\dagger + a_u)\kappa} | \rho_0 \rangle \rangle 
\end{align*}
As $|\rho_0 \rangle \rangle$ is the ground-state of two copies of the critical Ising model, it is symmetric under both $Z_2$ Ising symmetries: this implies that $|\rho_0 \rangle \rangle$ satisfies $\kappa |\rho_0 \rangle \rangle  = |\rho_0 \rangle \rangle$. The Klein factor also squares to one, which lends to the identity $[ (c_j^\dagger +c_j)(a_j^\dagger + a_j) \kappa]^2 = 1$. From this, the action of the operator $\hat{\varepsilon}$ on $|\rho_0 \rangle \rangle$ can be seen to be

\beqn
    e^{g \sum \limits_u  (c_u^\dagger + c_u) (a_u^\dagger + a_u) \kappa}  |\rho_0 \rangle \rangle &=& \prod_u [\cosh(g) + \sinh(g) (c_u^\dagger + c_u)(a_u^\dagger +a_u)]|\rho_0 \rangle \rangle, \\
    \Rightarrow \langle \langle \rho_0 | e^{g \sum \limits_u  (c_u^\dagger + c_u) (a_u^\dagger + a_u) \kappa }  &=& \langle \langle \rho_0|\prod_u [\cosh(g) - \sinh(g) (c_u^\dagger + c_u)(a_u^\dagger +a_u)].
\eeqn
The numerator of the spin-spin correlation function becomes the following product:
\begin{align*}
     &\langle \langle \rho^D |\sigma^z_v \sigma^z_{v'} | \rho^D \rangle \rangle \\&= \langle \langle \rho_0 |\prod_u [\cosh(g) - \sinh(g) (c_u^\dagger + c_u)(a_u^\dagger +a_u)]  (c_v^\dagger +c_v) e^{i \pi \sum \limits_{v \leq w < v'} c_w^\dagger c_w} (c_{v'}^\dagger +c_{v'}) \prod_u [\cosh(g) + \sinh(g) (c_u^\dagger + c_u)(a_u^\dagger +a_u)]|\rho_0 \rangle \rangle\\
     &= \cosh(2g)^{N - |v-v'|-1} \\
    \times &\langle \langle \rho_0 |\prod_{v \leq u \leq v' } [\cosh(g) - \sinh(g) (c_u^\dagger + c_u)(a_u^\dagger +a_u)]  (c_v^\dagger +c_v) e^{i \pi \sum \limits_{v \leq w < v'} c_w^\dagger c_w} (c_{v'}^\dagger +c_{v'}) \prod_{v \leq u \leq v'} [\cosh(g) + \sinh(g) (c_u^\dagger + c_u)(a_u^\dagger +a_u)]|\rho_0 \rangle \rangle
\end{align*}
In the last line, for $u>v'$ and $u<v$, every factor of $[\cosh(g) - \sinh(g)(c_u^\dagger + c_u)(a_u^\dagger+a_u)]$ on the left has been commuted all the way to the right in the expectation value. The $\cosh(2g)$ prefactor comes from $[\cosh(g) - \sinh(g)(c_u^\dagger + c_u)(a_u^\dagger +a_u)][\cosh(g) + \sinh(g)(c_u^\dagger + c_u)(a_u^\dagger +a_u)] = \cosh^2(g)+\sinh^2(g) = \cosh(2g)$. Another factor of $\cosh(2g)$ is attained by performing this maneuver for $[\cosh(g)-\sinh(g)(c_v^\dagger+c_v)(a_v^\dagger+a_v)]$. However, this is not the case for the remaining terms. We see that 

\begin{align*}
    &[\cosh(g) - \sinh(g) (c_{v'}^\dagger + c_{v'})(a_{v'}^\dagger +a_{v'})] (c_v^\dagger +c_v) e^{i \pi \sum \limits_{v \leq w < v'} c_w^\dagger c_w} (c_{v'}^\dagger +c_{v'}) 
    \\&=   (c_v^\dagger +c_v) e^{i \pi \sum \limits_{v \leq w < v'} c_w^\dagger c_w} (c_{v'}^\dagger +c_{v'}) [\cosh(g) + \sinh(g) (c_{v'}^\dagger + c_{v'})(a_{v'}^\dagger +a_{v'})]
\end{align*}
and lastly for $v < u < v'$:
\begin{align*}
    &[\cosh(g) - \sinh(g) (c_{u}^\dagger + c_{u})(a_{u}^\dagger +a_{u})] (c_v^\dagger +c_v) e^{i \pi \sum \limits_{v \leq w < v'} c_w^\dagger c_w} (c_{v'}^\dagger +c_{v'}) \\
    &=   (c_v^\dagger +c_v) e^{i \pi \sum \limits_{v \leq w < v'} c_w^\dagger c_w} (c_{v'}^\dagger +c_{v'}) [\cosh(g) + \sinh(g) (c_{u}^\dagger + c_{u})(a_{u}^\dagger +a_{u})].
\end{align*}
The numerator of the correlation function reduces to 

\beqn
    &= \cosh(2g)^{N - |v-v'|} 
   \langle \langle \rho_0| (c_v^\dagger +c_v) e^{i \pi \sum \limits_{v \leq k < v'} c_k^\dagger c_k} (c_{v'}^\dagger +c_{v'}) \prod \limits_{v < u \leq v'} [1+2 \sinh(g) \cosh(g) (c_u^\dagger + c_u)(a_u^\dagger +a_u)]|\rho_0 \rangle \rangle.
\eeqn

In terms of fermions $a,c$, the Ising model is a free theory and Wick's theorem holds. This is true even though the Hamiltonian is not diagonal in this basis, of course. All the terms with an odd number of $a$'s must vanish by symmetry, and every remaining term except one will involve products of contractions two $a_j^\dagger+a_j$ fermions, i.e. products of the Majorana fermion two-point functions $\langle \langle \rho_0 |(a_u^\dagger+a_u)(a_{u}^\dagger+a_{u'})|\rho_0 \rangle \rangle$. This correlation function necessarily vanishes in the critical Ising model when $u \neq u'$. This is a standard identity (see e.g. Ref.~\cite{IsingModelBook}), but one can verify this explicitly by diagonalizing the Bogoliubov-de-Gennes form of the Ising model Hamiltonian and computing $\langle \psi|(a_u^\dagger+a_u)(a_{u}^\dagger+a_{u'})|\psi \rangle$ in terms of the resulting Majoranas. There is only one non-vanishing contribution yielding

\beqn
\langle \sigma^z_v \sigma^z_{v'} \rangle &=& \cosh(2g)^{-|v-v'|} \langle \langle \rho_0| (c_v^\dagger +c_v) e^{i \pi \sum \limits_{v \leq k < v'} c_k^\dagger c_k} (c_{v'}^\dagger +c_{v'}) | \rho_0 \rangle \rangle, \cr\cr
&=& \cosh(2g)^{-|v-v'|} \langle \langle \rho_0 | \sigma^z_v \sigma^z_{v'} |\rho_0 \rangle \rangle, \cr\cr
&\sim& \frac{\cosh(2g)^{-|v-v'|}}{|v-v'|^{1/4}}.
\eeqn
Due to the $\cosh(g)^{-|v-v'|}$ term this is \textit{short-ranged}. Of course, the spin-spin correlation function on the other layer $\langle \tilde{\sigma}^z_v \tilde{\sigma}^z_{v'} \rangle$ must take the same form. One can see that Eq.~\ref{fdecstateapp} is invariant under Kramers-Wannier duality implying that this is also true for $\langle \mu^z_v \mu^z_{v'} \rangle$ and $\langle \tilde{\mu}^z_v \tilde{\mu}^z_{v'} \rangle$.

\section{Appendix C: Calculation of the dimension of BCCOs $\hat{\mathcal{B}}_x(\alpha,\beta)$ in the compact boson}

In this section, we compute the scaling dimension of the BCCO $\hat{\mathcal{B}}_x(\alpha, \beta)$. This calculation can also be found in the appendix of Ref.~\cite{MasakiDefect}, but we include it to be self-contained. The two-point function of the BCCO is

\begin{align}
    \bigg \langle \hat{\mathcal{B}}_{x=0}(\beta, \alpha) \hat{\mathcal{B}}_{x = \ell}( \alpha, \beta)  \bigg \rangle &= \int \limits_{\substack{\phi(x \in [0,\ell],\tau = 0) = \alpha \\ \phi(x \not \in [0,\ell],\tau = 0) = \beta}} \mathcal{D} \phi \text{ } e^{- \frac{1}{2 \pi} \int \dd[2]{x} \text{ }(\partial \phi)^2} \bigg / \int \limits_{\phi(x ,\tau = 0) = \beta } \mathcal{D} \phi \text{ } e^{- \frac{1}{2 \pi} \int \dd[2]{x} \text{ }(\partial \phi)^2},\\
    &= \int \limits_{\substack{\phi(x \in [0,\ell],\tau = 0) = \alpha - \beta \\ \phi(x \not \in [0,\ell],\tau = 0) = 0}} \mathcal{D} \phi \text{ } e^{- \frac{1}{2 \pi} \int \dd[2]{x} \text{ } (\partial \phi)^2} \bigg / \int \limits_{\phi(x ,\tau = 0) = 0 } \mathcal{D} \phi \text{ } e^{- \frac{1}{2 \pi} \int \dd[2]{x} \text{ }(\partial \phi)^2}.
\end{align}
In the second line, we've exploited the $\U(1)$ symmetry of the compact boson to shift $\phi \rightarrow \phi - \beta$ so that the correlation function only depends on the difference of phases $  \alpha - \beta$. The compact boson now satisfies the following boundary conditions:
\begin{align}
&\phi(x,0) = 
\begin{cases}
    \alpha - \beta & x \in [0,\ell]\\
    0 & x \not \in [0,\ell]
\end{cases} \label{Bosonbc}\\
&\phi(x,\tau \rightarrow \infty) = 0.
\end{align}
We decompose the field into two pieces $\phi(x,\tau) = \tilde{\phi}(x,\tau)+\phi_{\text{cl},n}(x,\tau)$. The $\tilde{\phi}(x,\tau)$ is fluctuating compact boson with uniform boundary conditions $\phi(x,0) = 0, \phi(x,\tau \rightarrow \infty) = 0$ while $\phi_{\text{cl},n}$ obeys Eq.~\ref{Bosonbc} and the equation of motion, $\partial^2 \phi_{\text{cl},n} = 0$. Since $\phi$ is compact, the path integral must be invariant under shifting of the boundary conditions by $2 \pi$. To incorporate this periodicity, different windings of the boundary conditions must be summed over. This is achieved by the solution
\beqn
\phi_{\text{cl},n}(x,\tau) = \frac{\alpha - \beta + 2\pi n}{\pi} \bigg ( \arctan \frac{\tau}{x - \ell} - \arctan \frac{\tau}{x} \bigg ).
\eeqn
where $n$ is summed over all integers in the path integral. With this, the BCCO correlation function reduces to 

\beqn
  \bigg \langle \hat{\mathcal{B}}_{x = 0}(\beta, \alpha) \hat{\mathcal{B}}_{x = \ell}( \alpha, \beta)  \bigg \rangle &=& \sum_{n = -\infty}^\infty e^{-\frac{1}{2 \pi} \int \dd[2]{x} \text{ } (\partial \phi_{\text{cl},n})^2} \int \limits_{\phi(x ,\tau = 0) = 0} \mathcal{D} \tilde{\phi} \text{ } e^{- \frac{1}{2 \pi} \int \dd[2]{x} \text{ } (\partial \tilde{\phi})^2  + 2 \partial \tilde{\phi} \cdot \partial \phi_{\text{cl},n}} \bigg / \int \limits_{\phi(x ,\tau = 0) = 0 } \mathcal{D} \phi \text{ } e^{- \frac{1}{2 \pi} \int \dd[2]{x} \text{ }(\partial \phi)^2},\\
 &=& \sum_{n = -\infty}^\infty e^{-\frac{1}{2 \pi} \int \dd[2]{x} \text{ } (\partial \phi_{\text{cl},n})^2} \label{BCCOsum}.
\eeqn
The cross term in the exponential vanishes from integrating by parts and using the fact that $\phi_{\text{cl},n}$ solves the equation of motion. The integral $\frac{1}{2 \pi} \int \dd[2]{x} (\partial \phi_\text{cl,n})^2$ can be computed using a UV cutoff $a$ as $\frac{(\alpha - \beta + 2 \pi n)^2}{\pi^2} \log \frac{\ell}{a}$. Hence at large $\ell$, this correlation function scales as $\langle \hat{\mathcal{B}}_{x = 0}(\beta, \alpha)  \hat{\mathcal{B}}_{x = \ell}(  \alpha,\beta)  \rangle \sim \ell^{-2 \Delta_{\hat{\mathcal{B}}}}$, where $\Delta_{\hat{\mathcal{B}}}$ is the value of $\frac{(\alpha - \beta + 2\pi n)^2}{\pi^2}$ that dominates the sum in in Eq.~\ref{BCCOsum}, that is

\beqn
\Delta_{\hat{\mathcal{B}}} = \min_{n \in \mathbb{Z}}\frac{1}{2} \bigg ( \frac{\alpha - \beta + 2\pi n }{\pi} \bigg )^2  = \frac{1}{2} \bigg ( \frac{\Delta \phi}{\pi} \bigg )^2,
\eeqn
where $\Delta \phi \leq \pi $ is the smallest distance between $\alpha$ and $\beta$ on the circle, $\Delta \phi = \min \{(\alpha - \beta) \mod{2 \pi}, 2 \pi -  (\alpha - \beta) \mod {2 \pi}\}$.

\section{Appendix D: Majorana zero modes, $e$-kinks, $m$-kinks, and fusion}

In the $em$-kink problem, the BCCOs $\hat{\mathcal{B}}_x(0, \frac{\pi}{2}),\hat{\mathcal{B}}_x(\frac{\pi}{2}, 0)$ have scaling dimension and spin $\Delta_{\hat{\mathcal{B}}} = s =\frac{1}{8}$, and are analogous to a pair of two distinct Majorana zero modes. This interpretation is simplest to make in the original inter-layer defect problem in two layers of the Ising CFT. The defect is $\sim -g' \int_{0}^{L/2} \dd{x} \sigma(x,0) \tilde{\sigma}(x,0) - g' \int_{-L/2}^{0} \mu(x,0) \tilde{\mu}(x,0)$, which has the following two effects upon flowing to the IR

\medskip
\textit{(1): } The $L/2>x>0$ piece of the defect $\sim -g' \int_{0}^{L/2} \dd{x} \sigma(x,0) \tilde{\sigma}(x,0)$ pins the spins in both layers along the defect in either the $\uparrow$, or $\downarrow$ directions.

\textit{(2): } The $-L/2<x<0 $ piece of the defect $\sim -g' \int_{0}^{L/2} \dd{x} \mu(x,0) \tilde{\mu}(x,0)$ pins the disorder parameters, i.e. the dual spins, in both layers along the defect in either the $\uparrow$, or $\downarrow$ directions.

\medskip
These are relevant defects, and end up cutting the connection between both sides of each layer. There are three conformal boundary conditions in the Ising CFT in terms of the original spins: the ``fixed" or Dirichlet boundary conditions $(\uparrow),(\downarrow)$ and the ``free" or Neumann boundary condition $(F)$. Along the cut, for $L/2>x>0$, the spins are ordered and hence obey either the $(\uparrow)$ or $(\downarrow)$ boundary condition in both layers. From the Kramers-Wannier duality, the ordering of the disorder parameters forces the original spins to be disordered. This can be thought of as enforcing the $(F)$ free/Neumann boundary condition on the original spins along the $-L/2 < x < 0$ region of the cut. \textit{The $\frac{\pi}{2}$ domain walls on the boundary of the compact boson now have the following interpretation as transitions between fixed and free boundary conditions in two layers of the Ising CFT: $(\uparrow \uparrow) \rightarrow (FF), (\downarrow \downarrow) \rightarrow (FF)$ or vice-versa.} For the BCCOs, this means that $\hat{\mathcal{B}} (0, \frac{\pi}{2} )$ and $\hat{\mathcal{B}} ( \frac{\pi}{2}, 0 )$ are equivalent to BCCOs in the two-layered Ising CFT $\hat{\mathcal{B}}_x(\uparrow \uparrow, FF),\hat{\mathcal{B}}_x(\downarrow \downarrow, FF)$ and $\hat{\mathcal{B}}_x(FF, \uparrow \uparrow),\hat{\mathcal{B}}_x(FF, \downarrow \downarrow)$ respectively. These are products of BCCOs that exchange free and fixed boundary conditions in a single Ising CFT, which have dimension and spin $\Delta = s = \frac{1}{16}$ \cite{cardybCFT1,LewellenBCCO}. The gapless boundary of Ising topological order is one of the two chiral halves of the Ising CFT, for which Majorana zero modes correspond to non-local twist fields with dimension and spin $\Delta = s = \frac{1}{16}$. The CFT fusion rules allow the twist field to fuse with one another into either the vacuum or a Majorana fermion (with dimension and spin $\Delta = s = 1/2$). These free-to-fixed (and vice-versa) BCCOs of the Ising CFT have the same dimension and spin as these twist fields, and obey similar CFT fusion rules in that they can fuse into the vacuum or a Majorana fermion boundary operator \cite{LewellenBCCO}. In this sense, the operators $\hat{\mathcal{B}}_x(0, \frac{\pi}{2}), \hat{\mathcal{B}}_x( \frac{\pi}{2},0)$ may be regarded analogously to pairs of these objects that are forced to live on the boundary of the CFT. Since a single $em$-kink in the decoherence problem is mapped to two $\frac{\pi}{2}$ domain walls, it is equivalent to \textit{four} of these analogous Majorana zero modes.

As a technical clarification, on the cylinder and torus, two layers of the Ising CFT are dual to a $Z_2$ orbifold boson (i.e. the compact boson with extra identification $\phi \sim - \phi$). The orbifold boson has actually two distinct $\phi = 0$ Dirichlet boundary conditions, which correspond to the $(\uparrow \uparrow)$ and $(\downarrow \downarrow )$ boundary conditions in the Ising CFT. Consequently, there are actually two distinct $\frac{\pi}{2}$ domain wall BCCOs for the orbifold boson which again correspond to the BCCOs from $(\uparrow \uparrow \rightarrow FF)$ and $(\downarrow \downarrow \rightarrow FF)$ and vice-versa, so that the counting of conformal boundary conditions is consistent between the two theories. We refer the reader to Ref.~\cite{MasakiDefect} for more details. The orbifolding of the boson does not change the $\log (L/2)$ scaling of $\cF_{em \text{ kink}}$ and $\Delta_{\hat{\mathcal{B}}}$ presented in the main text.

\bigskip

\begin{figure}[t]
    \centering
    \includegraphics[width=1\textwidth]{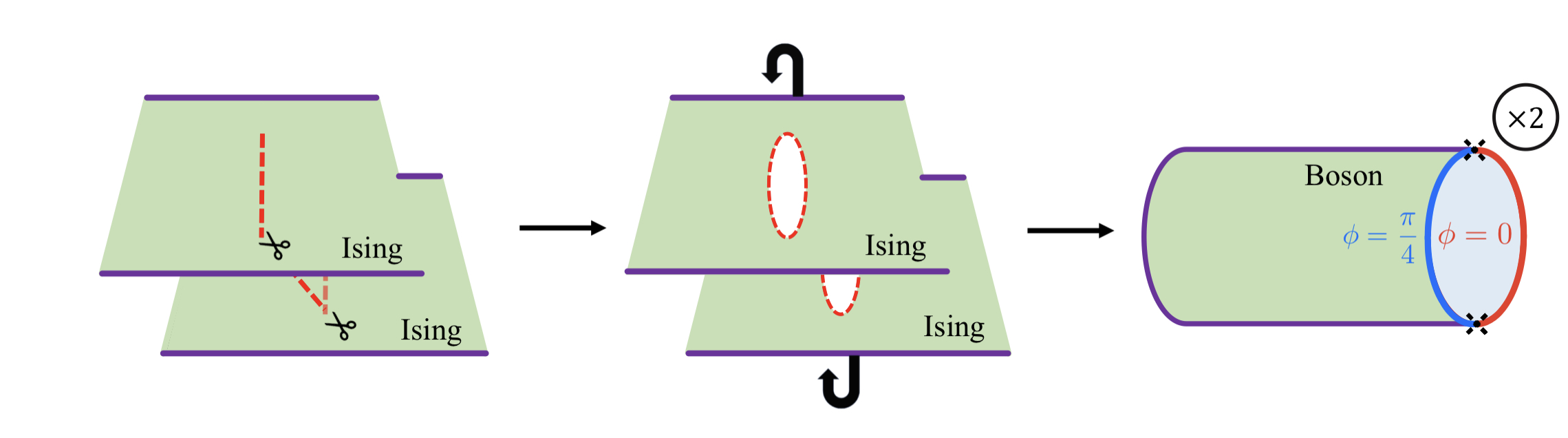}
    \caption{The inter-layer defect on half the line partially cuts each layer of the Ising CFT. Each layer can separately be folded over its respective cut, and after bosonizing each folded layer separately one arrives as two copies of the cylinder partition function of the compact boson with a domain wall between the $\phi = 0$ and $\phi = \frac{\pi}{4}$ boundary conditions.}
    \label{eKinksew}
\end{figure}

Naturally, there are other kinds of geometries of decoherence one can consider. The simplest possible other two examples involve decohering only half the critical boundary of toric code, i.e. we allow $e$ anyon tunneling or $m$ anyon tunneling for only $0 < v < L/2$. We refer to these as the $e$-kink and $m$-kink respectively. The $em$-kink is simply the fusion of an $e$-kink and an $m$-kink, which in $\tr{(\rho^D)^2}$, will correspond to the fusion of domain walls (BCCOs) in the compact boson. The decoherence channel for the $e$-kink is

\beqn
\varepsilon_v(\rho_0) \sim 
\begin{cases}
    (1-p)\rho_0 + p \sigma^z_v \rho_0 \sigma^z_v & L/2>v>0\\
    \rho_0 & -L/2<v<0
\end{cases}. \label{ekinkDec}
\eeqn
The decoherence channel for the $m$-kink is merely the Kramers-Wannier dual of Eq.~\ref{ekinkDec}. $\cZ_2$ under this decoherence channel is mapped to two layers of the Ising CFT with the following inter-layer defect $\sim -g' \int_0^{L/2} \dd{x} \sigma(x,0) \tilde{\sigma}(x,0)$. This defect only partially cuts the theory along $0 < x < L/2$, where the spins along the cuts in both layer are forced to point in the same direction. One can make use of Cardy's ``folding trick" to fold each layer of the Ising CFT in half over its cut. Bosonizing each folded layer individually leads to, once again, two cylinder partition functions of a compact boson with a domain wall on the cylinder's boundary. We illustrate this sewing procedure in Fig.~\ref{eKinksew}. It was shown in Ref.~\cite{MasakiDefect} that folding the Ising CFT with \textit{no defect} and then bosonizing leads to a Dirichlet boundary condition of $\phi = \frac{\pi}{4}$ for the compact boson. This can be intuitively understood as if there is no defect, the folded theory must be invariant under Kramers-Wannier duality. The duality acts as $\phi \rightarrow \frac{\pi}{2} - \phi$, which leaves the $\phi = \frac{\pi}{4}$ Dirichlet boundary condition unchanged.
Each $\frac{\pi}{4}$ domain wall is created by an insertion of the BCCO $\hat{\mathcal{B}}_x(0, \frac{\pi}{4})$ or $\hat{\mathcal{B}}_x(0, \frac{\pi}{4})$. The scaling dimension and spin of these operators can be read off from Eq.~\ref{BCCOdim} to be $\Delta_{\hat{\mathcal{B}}} = s = \frac{1}{32}$, and so the $e$-kink free energy is $\cF_{e \text{ kink}} = - \frac{1}{8} \log (\frac{L}{2})$. The case for the $m$-kink is related to that of the $e$-kink by Kramers-Wannier duality. This results in $\cZ_2$ being mapped to two cylinder partition functions of the compact boson with $\frac{\pi}{4}$ domain walls on each cap, \textit{this time corresponding to the transition between $\phi = \frac{\pi}{4}$ and $\phi = \frac{\pi}{2}$ Dirichlet boundary conditions}. The dimension and spin of the inserted BCCOs, now being $\hat{\mathcal{B}}_x(\frac{\pi}{4}, \frac{\pi}{2}), \hat{\mathcal{B}}_x(\frac{\pi}{2}, \frac{\pi}{4})$ are again $\Delta_{\hat{\mathcal{B}}} = s = \frac{1}{32}$.

Two different boundary domain walls for the compact boson, and their corresponding BCCOs obey an intuitive fusion rule. A domain wall that interpolates between the $\phi = \alpha$ and $\phi = \beta$ Dirichlet boundary conditions is allowed to fuse with another that interpolates between $\phi = \beta'$ and $\phi = \gamma'$ only when $\beta = \beta'$. If so, they are allowed to fuse into the new domain wall that interpolates between $\phi = \alpha$ and $\phi = \gamma$. For example, one can fuse our two different $\frac{\pi}{2}$ domain walls: a domain wall from $\phi = 0$ to $\phi = \frac{\pi}{4}$ for the $e$-kink and a domain wall from $\phi = \frac{\pi}{4}$ to $\phi = \frac{\pi}{2}$ for the $m$ kink. The resulting domain wall is the $\frac{\pi}{2}$ domain wall from $\phi = 0$ to $\phi = \frac{\pi}{2}$, exactly matching those of the $em$-kink problem. In terms of the BCCOs for the compact boson, this is the statement that $\hat{\mathcal{B}}_x(0, \frac{\pi}{4}) \times \hat{\mathcal{B}}_x(\frac{\pi}{4}, \frac{\pi}{2}) \sim  \hat{\mathcal{B}}_x(0, \frac{\pi}{2})$ and likewise for the conjugate BCCO. In this sense, in the doubled Hilbert space, the $e$-kink and $m$-kink fuse to an $em$-kink.

\end{document}